\documentstyle[12pt,twoside,fleqn,espcrc1,epsfig]{article}

\newcommand{\AmS}{{\protect\the\textfont2
  A\kern-.1667em\lower.5ex\hbox{M}\kern-.125emS}}

\title{Baryons, their interactions and the chiral symmetry of QCD}

\author{L. Ya. Glozman\address{Institute for Theoretical Physics, 
        University of Graz,\\ Universit\"atsplatz 5, A-8010 Graz, Austria}}%

\begin{document}
\maketitle

\begin{abstract}
An implication of the spontaneous chiral symmetry breaking in QCD is that at 
low energy and resolution there appear quasiparticles - constituent quarks
and Goldstone bosons. Thus, light and strange baryons should be considered
as systems of three constituent quarks with confining interaction and
a chiral interaction that is mediated by Goldstone bosons between the
constituent quarks. We show how the flavor-spin structure and sign of
the short-range part of the Goldstone boson exchange interaction reduces the
$SU(6)_{FS}$ symmetry down to $SU(3)_F \times SU(2)_S$, induces hyperfine
splittings and provides correct ordering of the lowest states with positive
and negative parity. We present a unified description of 
light and strange baryon spectra calculated in a semirelativistic framework.  
It is demonstrated that the same 
short-range part of Goldstone boson exchange 
also induces strong short-range repulsion in $NN$ system when the
latter is treated as $6Q$ system. Thus all main ingredients of $NN$
interaction are implied by the chiral constituent quark model since the
long- and intermediate-range attraction appears in the present framework
due to pion and correlated two-pion exchanges between quarks belonging 
to different nucleons. We also find a very strong short-range repulsion
in $\Lambda\Lambda$ system with $J^P=0^+$. It then suggests that the
compact $H$-particle should not exist.
\end{abstract}

\section{Spontaneous Chiral Symmetry  Breaking and its Implications
for  Low-Energy QCD}

The QCD Lagrangian with three light flavors has a global symmetry

\begin{equation}SU(3)_{\rm L} \times SU(3)_{\rm R} \times U(1)_{\rm V} \times
U(1)_{\rm A}.  \label{1} \end{equation}

\noindent
 At low temperatures and densities the  
$SU(3)_{\rm L} \times SU(3)_{\rm R}$chiral symmetry is spontaneously
broken down to $SU(3)_{\rm V}$ in the QCD vacuum, i.e., realized in the hidden
Nambu-Goldstone mode. A direct evidence for the spontaneously broken
chiral symmetry is a nonzero value of the quark condensates for the
light flavors
$<|\bar{q}q|> \approx -(240-250 {\rm
MeV})^3,$
which represent the order parameter. That this is indeed so, we know from
three independent sources: current algebra,
QCD sum rules, and lattice gauge calculations.
There are two important generic consequences of the spontaneous chiral symmetry
 breaking. The first one is an appearance of the octet
of pseudoscalar mesons of low mass, $\pi, {\rm K}, \eta$, which represent
the associated approximate Goldstone bosons. The second one is that valence
quarks acquire a dynamical or constituent mass.
Both these
 consequences of the spontaneous chiral symmetry  breaking
are well illustrated by, e.g. the $\sigma$-model \cite{LEVY} or the Nambu and
Jona-Lasinio model \cite{Nambu}.
We cannot say at the
moment for sure what the microscopical reason for spontaneous chiral symmetry
 breaking in the QCD vacuum is. It was suggested that this occurs
when quarks propagate through instantons in the QCD vacuum \cite{DIP}.

For the low-energy baryon properties it is only essential that
beyond the spontaneous chiral symmetry breaking scale (i.e. at low resolution)
 new
dynamical degrees of freedom appear - constituent quarks and chiral
fields which couple together \cite{WEIN,MAG}. The low-energy baryon 
properties are mainly determined
by these dynamical degrees of freedom and the confining interaction.

There is a very good analogy in solid state physics. 
For instance, the fundamental degrees of freedom in crystals are ions
in the lattice, electrons and the electromagnetic field. Nevertheless, in order
to understand electric conductivity, heat capacity, etc. we instead work
with  "heavy electrons" with dynamical mass, phonons and their interaction.
In this case a complicated electromagnetic interaction of the electrons with
the ions in the lattice is "hidden" in the dynamical mass of the electron
and the
interactions among ions in the lattice are eventually responsible  for the
 collective excitations of the lattice - phonons,
which are Goldstone bosons of the spontaneously broken translational
invariance in the lattice of ions.
As a result, the theory becomes rather
simple - only the electron and phonon degrees of freedom and electron-phonon 
interactions
are essential for all the  properties of crystals mentioned above.

We have recently suggested \cite{GLO2} that in the low-energy regime 
light and strange baryons
should be considered as a system of 3 constituent quarks with an effective
$Q-Q$ interaction that is formed of central confining part and a chiral
interaction mediated by the Goldstone bosons between constituent quarks.
This physical picture allows to understand a structure of the baryon spectrum
and to solve, in particular,  the long-standing problem of ordering of the
lowest baryons with positive- and negative-parity.

\section{The Goldstone Boson Exchange Interaction 
and the Structure of  Light and Strange Baryon Spectra}

The
coupling of the constituent  quarks and the pseudoscalar Goldstone
bosons will (in the $SU(3)_{\rm F}$ symmetric approximation) have
the form $ig\bar\psi\gamma_5\vec\lambda^{\rm F}
\cdot \vec\phi\psi$ (or
$g/(2m)\bar\psi\gamma_\mu
\gamma_5\vec\lambda^{\rm F}
\cdot \psi \partial^\mu\vec\phi$).
  A coupling of this
form, in a nonrelativistic reduction for the constituent quark spinors,
will -- to lowest order -- give rise the 
$\sim\vec\sigma \cdot \vec q \vec\lambda_i^{\rm F}$ structure of the 
meson-quark vertex, where $\vec q$ is meson momentum. Thus, the structure
of the $Q-Q$ potential in momentum representation is

\begin{equation}V(\vec q) \sim
\vec\sigma_i\cdot\vec q \sigma_j\cdot\vec q ~
\vec\lambda_i^{\rm F}\cdot\vec\lambda_j^{\rm F}
D(q^2) F(q^2)
,\label{2} \end{equation}

\noindent
where $D(q^2)$ is dressed Green function for chiral field which includes
both nonlinear terms of chiral Lagrangian and fermion loops, $F(q^2)$ is
meson-quark formfactor which takes into account internal structure
of quasiparticles. At big distances ($\vec q \rightarrow 0$), 
$D(q^2) \rightarrow D_0({\vec q}^2)= -({\vec q}^2 + \mu^2)^{-1}$, where
$D_0({\vec q}^2)$ is free Klein-Gordon Green function in static
approximation. Thus, $D(\vec q = 0) \not= \infty$. It then follows from
(\ref{2}) that $V(\vec q = 0) = 0$, which is equivalent to 
$\int d\vec r V(\vec r) = 0$. Since at big interquark separations
the spin-spin component of the pseudoscalar-exchange interaction is

\begin{equation}
V(r_{ij}) \sim \vec\sigma_i\cdot\vec\sigma_j
\vec\lambda_i^{\rm F}\cdot\vec\lambda_j^{\rm F}\mu^2 
\frac {e^{-\mu r_{ij}}}{r_{ij}}, 
\end{equation}

\noindent
it then follows from the volume
integral constraint  that at short interquark separations the spin-spin
interaction should be opposite in sign as compared to the Yukawa tail and
very strong. {\it It is this short-range part of the Goldstone boson exchange
(GBE) interaction between constituent quarks that is of crucial importance
for baryons: it has a sign appropriate to reproduce the level splittings
and dominates over the Yukawa tail towards short distances.} In a
oversimplified consideration with a free Klein-Gordon Green function instead
of the dressed one in (\ref{2}) and with $F(q^2)=1$, one obtains the
following  spin-spin component of $Q-Q$ interaction:

\begin{equation}V(r_{ij})=
\frac{g^2}{4\pi}\frac{1}{3}\frac{1}{4m_im_j}
\vec\sigma_i\cdot\vec\sigma_j\vec\lambda_i^{\rm F}\cdot\vec\lambda_j^{\rm F}
\{\mu^2\frac{e^{-\mu r_{ij}}}{ r_{ij}}-4\pi\delta (\vec r_{ij})\}
.\label{3} \end{equation}

 Consider first, for the purposes of illustration, a schematic model
which neglects the radial dependence
of the potential function $V(r_{ij})$ in (\ref{3}), and assume a harmonic
confinement among quarks as well as $m_{\rm u}=m_{\rm d}=m_{\rm s}$.
In this model

\begin{equation}H_\chi = -\sum_{i<j}C_\chi~
\vec \lambda^{\rm F}_i \cdot \vec \lambda^{\rm F}_j\,
\vec
\sigma_i \cdot \vec \sigma_j.\label{4} \end{equation}

If the only interaction between the
quarks were the flavor- and spin-independent harmonic confining
interaction, the baryon spectrum would be organized in multiplets
of the symmetry group $SU(6)_{\rm FS} \times U(6)_{\rm conf}$. In this case
the baryon masses would be determined solely by the orbital structure,
and the spectrum would be organized in an {\it alternative sequence
of positive and negative parity states.}
The Hamiltonian (\ref{4}), within a first order perturbation theory,
 reduces the $SU(6)_{\rm FS} \times U(6)_{\rm conf}$ symmetry down to
 $SU(3)_{\rm F}\times SU(2)_{\rm S}\times U(6)_{\rm conf}$, which automatically
implies a splitting between the octet and decuplet baryons.

The  two-quark matrix elements of the
interaction (\ref{4}) are:

$$<[f_{ij}]_{\rm F}\times [f_{ij}]_{\rm S} : [f_{ij}]_{\rm FS}
{}~| -C_\chi\vec \lambda^{\rm F}_i \cdot \vec \lambda_j^{\rm F}
\vec\sigma_i \cdot \vec \sigma_j
{}~|~[f_{ij}]_{\rm F} \times [f_{ij}]_{\rm S} : [f_{ij}]_{\rm FS}> $$
\begin{equation}=\left\{\begin{array}{rl} -{4\over 3}C_\chi & [2]_{\rm
F},[2]_{\rm S}:[2]_{\rm FS} \\
-8C_\chi & [11]_{\rm F},[11]_{\rm S}:[2]_{\rm FS} \\
4C_\chi & [2]_{\rm F},[11]_{\rm S}:[11]_{\rm FS}\\ {8\over
3}C_\chi & [11]_{\rm F},[2]_{\rm S}:[11]_{\rm
FS}\end{array}\right..\label{5} \end{equation}

\noindent
{}From these the following important properties may be inferred:

(i) At short range the GBE
interaction  is attractive in the symmetric FS pairs and
repulsive in the antisymmetric ones.

(ii) Among the FS-symmetrical pairs,
the flavor antisymmetric pairs experience
a much larger attractive interaction than the flavor-symmetric
ones, and among the FS-antisymmetric pairs
the strength of the repulsion in flavor-antisymmetric
pairs is considerably weaker than in the symmetric ones.

Given these properties we conclude, that
with the given flavor symmetry, the more symmetrical the FS
Young pattern is for a baryon - the more attractive contribution at short
range comes from the GBE. For two identical
flavor-spin Young patterns $[f]_{\rm FS}$ the attractive contribution
at short range is larger for  the more antisymmetrical
flavor Young pattern $[f]_{\rm F}$.

The flavor-spin symmetry and the sign of the GBE interaction at short range
allow to  understand the structure of the low-lying baryon spectrum.

For the octet states ${\rm N}$, $\Lambda$, $\Sigma$,
$\Xi$ ($N=0$ shell, $N$ is the number of harmonic oscillator excitations
in a 3-quark state) as well as for their first
radial excitations of positive parity
 ${\rm N}(1440)$, $\Lambda(1600)$, $\Sigma(1660)$,
$\Xi(?)$ ($N=2$ shell) the flavor and spin symmetries
are $[3]_{\rm FS}[21]_{\rm F}[21]_{\rm S}$, and the contribution of the
Hamiltonian (\ref{4})
 is $-14C_\chi$. For the decuplet states
$\Delta$, $\Sigma(1385)$, $\Xi(1530)$, $\Omega$ ($N=0$ shell)
the flavor and spin symmetries,
as well as the corresponding matrix element, are
$[3]_{\rm FS}[3]_{\rm F}[3]_{\rm S}$
and $-4C_\chi$, respectively. The  negative parity excitations
($N=1$ shell) in the ${\rm N}$, $\Lambda$ and $\Sigma$ spectra 
(${\rm N}(1535)$ - ${\rm N}(1520)$, $\Lambda(1670)$ - $\Lambda(1690)$
and $\Sigma(1750)$ - $\Sigma(?)$)
are described by the $[21]_{\rm FS}[21]_{\rm F}[21]_{\rm S}$ symmetries, and
the contribution
of the interaction (\ref{4}) in this case is $-2C_\chi$. The first negative
parity excitation in the $\Lambda$ spectrum ($N=1$ shell)
$\Lambda(1405)$ - $\Lambda(1520)$ is flavor singlet $[21]_{\rm FS}[111]_{\rm
F}[21]_{\rm S}$,
and, in this case, the corresponding matrix element is $-8C_\chi$. The latter
state is unique and is absent in other spectra due to its flavor-singlet
nature.

These  matrix elements alone suffice to prove that
the ordering of the lowest positive and negative parity states
in the baryon spectrum will be correctly predicted by
the chiral boson exchange interaction (\ref{4}).
The constant $C_\chi$ may be determined from the
N$-\Delta$ splitting to be 29.3 MeV.
The oscillator
parameter $\hbar\omega$, which characterizes the
effective confining interaction,
may be determined as  one half of the mass differences between the
first excited
$\frac{1}{2}^+$ states and the ground states of the baryons,
which have the same flavor-spin, flavor and spin symmetries
(e.g. ${\rm N}(1440)$ - ${\rm N}$, $\Lambda(1600)$ - $\Lambda$, $\Sigma(1660)$
- $\Sigma$),
to be
$\hbar\omega \simeq 250$ MeV. Thus the two free parameters of this simple model
are fixed and we can  make now predictions.

In the ${\rm N}$ and $\Sigma$ sectors the mass
difference between the lowest
excited ${1\over 2}^+$ states (${\rm N}(1440)$, $\Lambda(1600)$, 
and $\Sigma(1660)$)
and ${1\over 2}^--{3\over 2}^-$ negative parity pairs
 (${\rm N}(1535)$ - ${\rm N}(1520)$, $\Lambda(1670)$ - $\Lambda(1690)$,
 and $\Sigma(1750)$ - $\Sigma(?)$, respectively) will then
be
\begin{equation}{\rm N},\Lambda,\Sigma:
\quad m({1\over 2}^+)-m({1\over 2}^--{3\over
2}^-)=250\, {\rm
MeV}-C_\chi(14-2)=-102\, {\rm MeV},\end{equation}
whereas for the lowest states in the $\Lambda$ system ($\Lambda(1600)$,
$\Lambda(1405)$ - $\Lambda(1520)$) it should be

\begin{equation}\Lambda:\quad m({1\over 2}^+)-m({1\over 2}^--{3\over
2}^-)=250\, {\rm
MeV}-C_\chi(14-8)=74\, {\rm MeV}.  \end{equation}

This simple example shows how the GBE interaction 
provides different ordering of the lowest positive and negative parity excited
states in the spectra of the nucleon and
the $\Lambda$-hyperon. This is a direct
consequence of the symmetry properties of the boson-exchange interaction
discussed at the beginning of this section.
Namely, the $[3]_{\rm FS}$ state in the ${\rm N}(1440)$
$\Lambda(1600)$and
$\Sigma(1660)$ positive parity resonances from the $N=2$ band feels a
much stronger
attractive interaction than the mixed symmetry state $[21]_{\rm FS}$ in the
${\rm N}(1535)$ - ${\rm N}(1520)$, $\Lambda(1670)$ - $\Lambda(1690)$
and $\Sigma(1750)$ -$\Sigma(?)$ resonances of negative parity ($N=1$ shell).
Consequently the masses of the
positive parity states ${\rm N}(1440)$, $\Lambda(1600)$  and
$\Sigma(1660)$ are shifted
down relative to the other ones, which explains the reversal of
the otherwise expected "normal ordering".
The situation is different for $\Lambda(1405)$ - $\Lambda(1520)$
and
$\Lambda(1600)$, as the flavor state of  $\Lambda(1405)$ - $\Lambda(1520)$ is
totally antisymmetric. Because of this the
$\Lambda(1405)$ - $\Lambda(1520)$ gains an
attractive energy, which is
comparable to that of the $\Lambda(1600)$, and thus the ordering
suggested by the confining oscillator interaction is maintained.

\section{Semirelativistic Chiral Constituent Quark Model}

In the semirelativistic chiral constituent quark model \cite{GPPVW,GPVW}
the dynamical part of the Hamiltonian consists of linear pairwise
confining interaction and the GBE interaction, which includes both
the short-range part (parametrized phenomenologically) and the long-range
Yukawa tail.  In the large $N_C$ limit
the axial anomaly vanishes \cite{WITT} and the $\eta'$ (the flavor singlet)
becomes the ninth Goldstone boson of $U(3)_R\times U(3)_L$ chiral symmetry
\cite{COLWITT}. Thus, the $\eta'$ should  be also taken into account,
but with coupling constant different as compared to octet Goldstone
bosons. The latter coupling constant can be fixed from the known pion-nucleon
coupling constant.

The kinetic-energy operator is taken in relativistic form,
$H_0 = \sum_{i=1}^3 \sqrt(p_i^2 + m_i^2)$. The semirelativistic 
three-quark Hamiltonian was solved along the stochastical variational
method \cite{VARGA} in momentum space. For the whole Q-Q potential the
model involves a total of 5 free parameters whose numerical values are
determined from the fit to all 34 confirmed low-lying states \cite{GPPVW,GPVW}. 
In fig. 1 we present the ground states as well as low-lying excited states
in $N$, $\Delta$, $\Lambda$, $\Sigma$, $\Xi$, and $\Omega$ spectra.
From the results of fig. 1 it becomes evident that within the chiral
constituent quark model a unified description of both nonstrange and
strange baryon spectra is achieved in good agreement with phenomenology.

It is instructive to learn how the GBE interaction affects the energy levels 
when it is switched on and its strength is gradually increased (Fig. 2).
Starting out from the case with confinement only, one observes that
the degeneracy of states is removed and the inversion of ordering
of positive and negative parity states 
is achieved in the $N$ spectrum, as well as for some states in the
$\Lambda$ spectrum, while the ordering of the lowest positive-negative
parity states is opposite in $N$ and $\Lambda$ spectra. The reason for
this behaviour is the flavor-spin structure and sign of the short-range
part of GBE, discussed in the previous section.

In the semirelativistic model we find that the confinement strength should
be consistent with Regge slopes and the string tension extracted in lattice
QCD. This is comfortable improvement over nonrelativistic model \cite{GPP}
where a much smaller value of the confinement strength must be chosen
in order to reproduce the low-lying spectrum.
\begin{figure}
\begin{center}
\epsfig{file=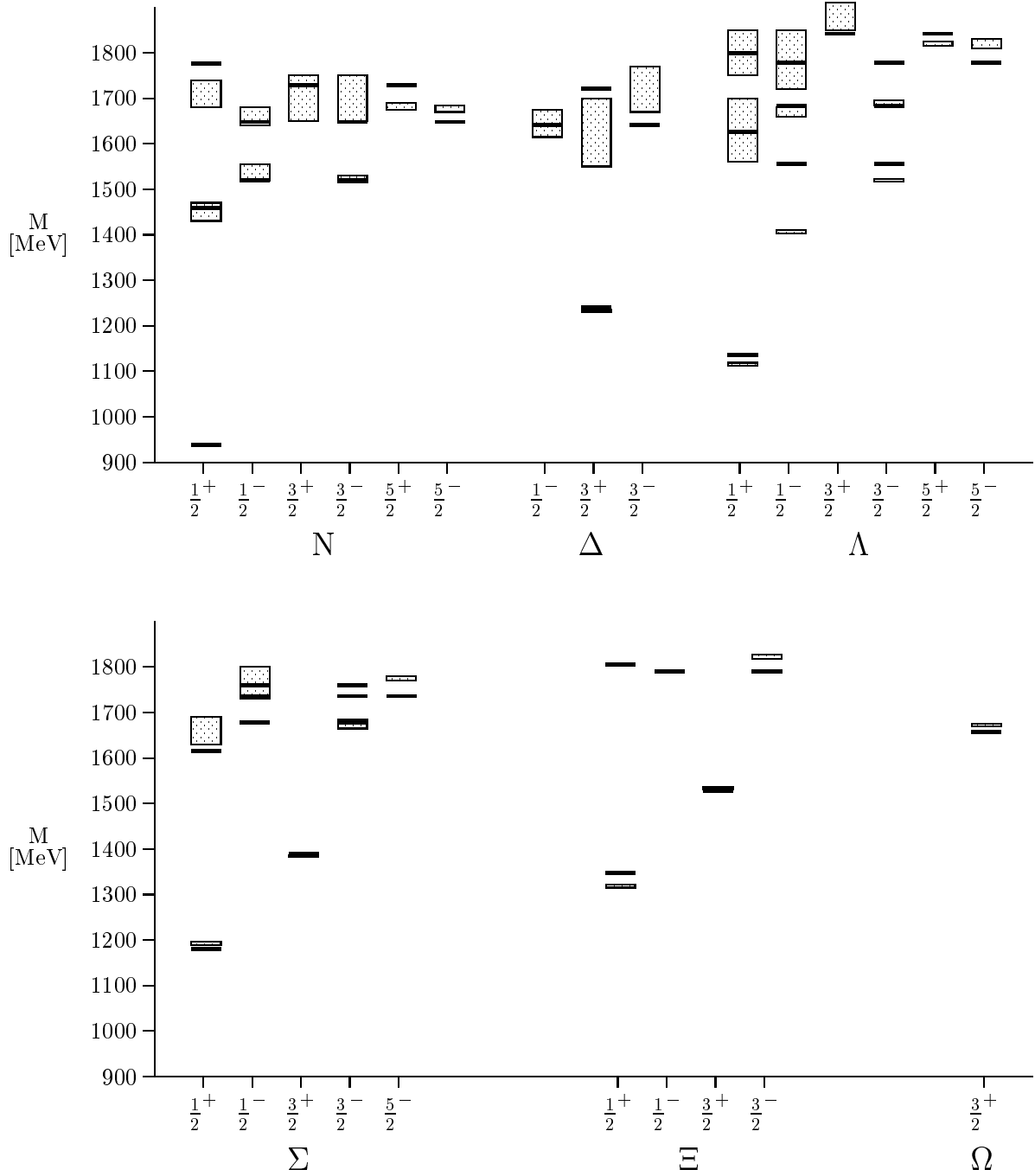}
\caption{Energy levels of the lowest light and strange baryon states 
(below 1850 MeV) with total
angular momentum and parity $J^P$. The shadowed boxes represent the experimental
values with their uncertainties.}
\end{center}
\end{figure}
\begin{figure}
\epsfig{file=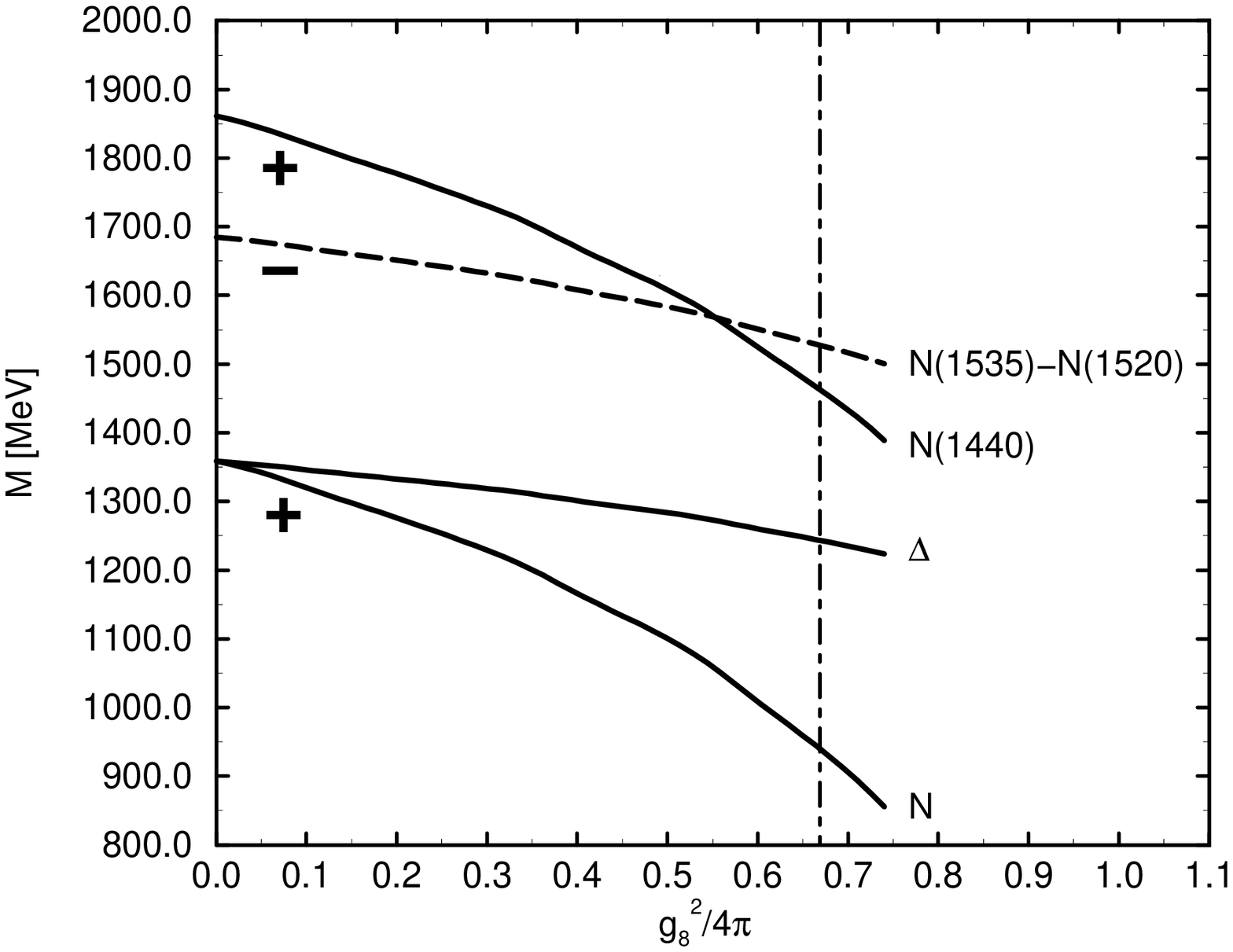,width=8cm}\hfill
\epsfig{file=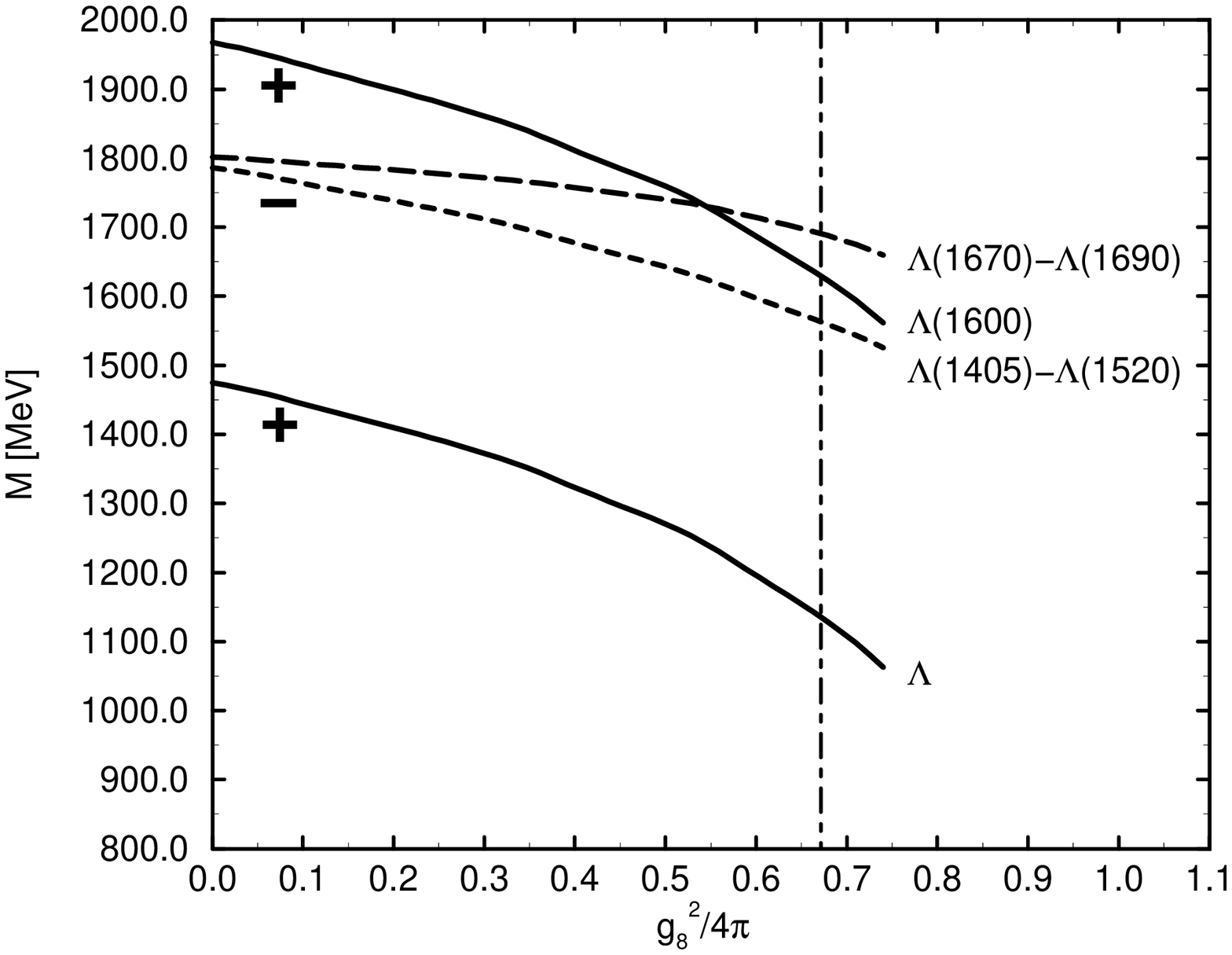,width=8cm}
\caption{Level shifts of some lowest baryons as a function of the 
strength of the GBE. Solid and dashed lines correspond to positive- 
and negative-parity states, respectively.}
\end{figure}
\begin{figure}
\begin{center}
\epsfig{file=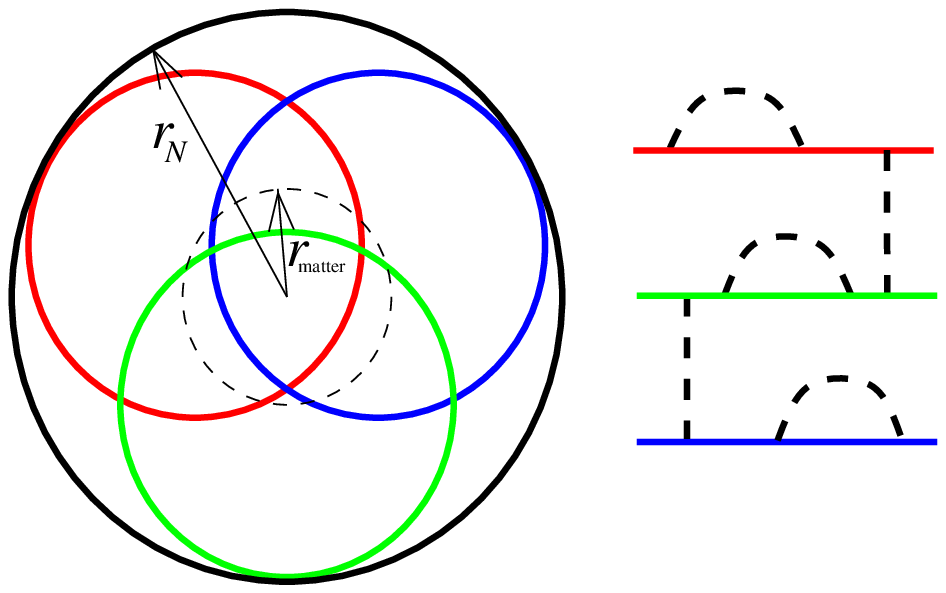}
\caption{Nucleon as it is seen in the low-energy and low-resolution regime.} 
\end{center}
\end{figure}

It is clear that the Fock components $QQQ\pi, QQQK, QQQ\eta$, and $QQQ\eta'$
(including meson continuum) cannot be completely integrated out in favour
of the meson-exchange $Q-Q$ potentials for some states above or near
the corresponding meson thresholds. Such a component in addition to the
main one $QQQ$ could explain e.g. an exceptionally big splitting of the
flavor singlet states $\Lambda(1405)-\Lambda(1520)$, since the $\Lambda(1405)$
lies below the $\bar K N$ threshold and can be presented as  $\bar K N$ bound
system \cite{DAL}. Note, that in the case of the present approach this old
idea is completely natural and does not contradict a flavor-singlet  $QQQ$ 
nature of $\Lambda(1405)$, while it would be in conflict with naive constituent
quark model where no room for mesons in baryons. An admixture of such components
will be  important in order to understand strong decays of some excited states.
While technically inclusion of such  components in addition to the main
one $QQQ$ in a coupled-channel approach is rather difficult task, it should
be considered as one of the most important future directions.

What is an intuitive picture of the nucleon in the low-energy regime? 
Nucleon consists of 3 constituent quarks which are very big objects due 
to their meson clouds (see fig. 3). These constituent quarks are all the time 
in strong overlap inside the nucleon. That is why the short-range part of 
GBE interaction (which is represented by the contact term in the oversimplified 
representation (\ref{3})) is so crucially important inside baryons.
When constituent quarks are well separated and there is a phase space for pion
propagation, the long-range Yukawa tail of GBE interaction as well as
correlated two-pion exchanges become very important. It is these parts
of meson exchange which produce the necessary long- and intermediate-range 
attraction in two-nucleon system.

\section{The Nucleon-Nucleon Interaction in a Chiral Constituent Quark Model}

So far, all studies of the short-range $NN$ interaction within the
constituent quark model were based on the one-gluon exchange interaction
between quarks. They explained the short-range repulsion in the $NN$ system
as due to the colour-magnetic part of OGE combined with quark interchanges
between 3Q clusters \cite{OKA}. It has been shown, however, that there
is practically no room for colour-magnetic interaction in light 
baryon spectroscopy and any appreciable amount of colour-magnetic interaction, 
in addition to GBE, destroys the spectrum \cite{GPPVW}. This conclusion
is confirmed by recent lattice QCD calculations \cite{LIU}. If so, the
question arises which interquark interaction is responsible for the
short-range $NN$ repulsion. Below I show that the same short-range part
of GBE which causes $N-\Delta$ splitting and produces good baryon spectra,
also induces a short-range repulsion in $NN$ system when the latter is
treated as 6Q system \cite{STPEPGL}.

In order to have a qualitative insight into the short-range $NN$ interaction
it is convenient to use an adiabatic Born-Oppenheimer approximation
for the internucleon potential:

\begin{equation} V_{NN}(R) = <H>_R - <H>_\infty, \label{6} \end{equation}

\noindent
where $\vec R$ is a collective coordinate which is the separation
distance between the two $s^3$ nucleons, $<H>_R$ is the lowest
expectation value of the 6Q Hamiltonian at fixed $R$, and $<H>_\infty$
is mass of two well-separated nucleons ($2m_N$) calculated with the same
Hamiltonian.

It is well known that when the separation $R$ approaches 0, then for both
$^3S_1$ and $^1S_0$ partial waves in the $NN$ system only two types
of 6Q configurations survive \cite{HARVEY}: 
$|s^6 [6]_O>$ and  $|s^4p^2 [42]_O>$, where $[f]_O$ is Young diagram,
describing spatial permutational symmetry in 6Q system. There are
a few different flavor-spin symmetries, compatible with spatial symmetries
above: $[6]_O[33]_{FS}$, $[42]_O[33]_{FS}$, $[42]_O[51]_{FS}$, 
$[42]_O[411]_{FS}$, $[42]_O[321]_{FS}$, and $[42]_O[2211]_{FS}$. Thus,
in order to evaluate the $NN$ interaction at zero separation between
nucleons (i.e. their short-range interaction) it is necessary to
diagonalize a $6Q$ Hamiltonian in the basis above and use the procedure (\ref{6}).
We adjust parameters of the "GBE interaction" and "confinement" in a 
nonrelativistic 3-body calculation  to reproduce low-lying $N$ and $\Delta$
spectrum \cite{GPP}. Clearly, that the nonrelativistic parametrization 
is only effective 
(relative to parameters exctracted from the semirelativistic fit).
Since we are interested at the moment only in qualitative effects, 
related to flavor-spin structure and sign of short-range GBE interaction, 
such an approach can be considered as reasonable. 

From the adiabatic Born-Oppenheimer approximation (\ref{6}) we find
that $V_{NN}(R=0)$ is highly repulsive in both $^3S_1$ and $^1S_0$
partial waves, with the core being of order 1 GeV. 
This repulsion implies a strong suppression of the $NN$ wave function 
in the nucleon overlap region.

Due to the specific flavor-spin symmetry of GBE the configuration
$[42]_O[51]_{FS}$ becomes highly dominant among other possible 6Q 
configurations at zero separation between nucleons (however, the "energy"
of this configuration is much higher than the energy of two well-separated
nucleons). The symmetry structure of this
dominant configuration induces "an additional" effective repulsion,
related to the "Pauli forbidden state" in this case \cite{NST}. As a result,
the s-wave $NN$ relative motion wave function has a node at short range.
The existence of a strong repulsion, related to the energy ballance in the
adiabatic approximation, suggests, however, that the amplitude of the
oscillating $NN$ wave function at short range will be strongly suppressed. 

Thus, within the chiral constituent quark model one has all the necessary
ingredients to understand microscopically the $NN$ interaction. There
appears strong short-range repulsion from the same short-range part of GBE
which also produces hyperfine splittings in baryon spectroscopy.
The long- and intermediate-range attraction in the $NN$ system is
automatically implied by the Yukawa part of pion-exchange and correlated
two-pion exchanges between quarks belonging to different nucleons.
With this first encouraging result, it might be worthwhile to perform
a more elaborate calculation of $NN$ and other baryon-baryon systems
within the present framework.

\section{Does H-particle exist?}

It is also interesting to see what will happen at short range
in other baryon-baryon systems. In particular, assuming that the
colour-magnetic interaction is a main reason for hyperfine splittings
in baryons, Jaffe has predicted that there should be well bound dibaryon
in $\Lambda\Lambda$ system with $J^P=0^+$, called H-particle \cite{JAFFE}.
In this specific case the  colour-magnetic interaction in 6q system becomes
the most attractive and implies an existence of well-bound dibaryon, stable
against $\Lambda\Lambda$ strong decay, since it should be well below this
threshold. This H-particle should be compact object in contrast to 
molecule-like state - deuteron.

It turns out, however, 
that within our approach there is very strong repulsion at short
range in  $\Lambda\Lambda$ system with $J^P=0^+$, coming from the same
short-range part of GBE, which induces hyperfine splittings and short-range
$NN$ repulsion. This short-range repulsion in $\Lambda\Lambda$ system 
can be obtained in a similar way \cite{H}  
to what has been done for $NN$ system in the previous section. To be more
specific, the lowest possible 6Q configuration in this case is
$|s^6 [6]_O [33]_{FS}>$, which lies  about 1 GeV above the 
$\Lambda\Lambda$ threshold. It then suggests, that in our approach to
baryon structure the compact H-particle should not exist. 

However, there is an attraction in $\Lambda\Lambda$ system at 
intermediate and long
distances which has the same nature like in two-nucleon system. 
The $\Lambda\Lambda$ system with $J^P=0^+$
is similar in this sense to $^1S_0$  NN partial wave, where the long-
and intermediate-range attraction is not enough to bind two-nucleon system.

 An extensive
experimental search for H-particle for 20 years has  showed no positive
results \cite{CHRIEN}.

\bigskip
\noindent
{\bf Acknowledgement}

It is my pleasure to thank D.O.Riska, Z.Papp, W.Plessas, K.Varga, R. Wagenbrunn,
Fl. Stancu, and S. Pepin, in collaboration with whom different results
discussed in this talk have been obtained.

\end{document}